\documentclass[twocolumn,prb,amsmath,amssymb,superscriptaddress]{revtex4}
\usepackage{graphicx}
\usepackage{bm}
\usepackage{bbm}
\usepackage{mathbbol}

\DeclareMathAlphabet{\mathpzc}{OT1}{pzc}{m}{it}

\begin{document}
\title{Two-spin dephasing by electron-phonon interaction in semiconductor double quantum dots}
\author{Xuedong \surname{Hu}}
\affiliation{Department of Physics, University at Buffalo, SUNY, Buffalo, NY 14260-1500}
\date{\today}
\begin{abstract}
We study electron-phonon interaction induced decoherence between two-electron singlet and triplet states in a semiconductor double quantum dot
using a spin-boson model.  We investigate the onset and time evolution of this dephasing, and study its dependence on quantum dot parameters
such as dot size and double dot separations, as well as the host materials (GaAs and Si). At the short time limit, electron-phonon interaction
only causes an incomplete initial Gaussian decay of the off-diagonal density matrix element in the singlet-triplet Hilbert space, a complete
long-time exponential decay due to phonon relaxation would eventually dominate over two-spin decoherence.  We analyze two-spin decoherence in
both symmetric and biased double quantum dots, identifying their difference in electron-phonon coupling and the relevant consequences.
\end{abstract}
 \maketitle

\section{Introduction}

Significant experimental progresses in the study of semiconductor spin qubits in the past few years\cite{Petta_Science05, Koppens_PRL08,
Shaji_NP08, Lansbergen_NP08, Tarucha_NP08, Amasha_PRL08, Vink_NP09, Foletti_NP09, Shin_PRL10, Petta_Science10, Xiao_PRL10, Reilly_PRL10,
Bluhm_NP11, Abe_PRB10, Morello_Nature10, Kouwenhoven_Nature10, Morton_Nature11} have reconfirmed the confined electron spins as one of the
leading candidates for the building block of a solid state quantum information processor. A decade of theoretical studies has mostly clarified
single spin decoherence channels and their relative importance in semiconductor quantum dot (QD) and donor confined
electrons,\cite{Khaetskii_PRB01, Merkulov_PRB02, Erlingsson_PRB02, Tahan_PRB02, Khaetskii_PRL02, deSousa_PRB03a, deSousa_PRB03b, Golovach_PRL04,
Coish_PRB04, Tahan_PRB05, Witzel_PRB05, Deng_PRB06, Witzel_PRB06, Yao_PRB06, Zhang_PRB06, Liu_NJP07, Witzel_PRB07, Deng_PRB08, Cywinski_PRL09,
Cywinski_PRB09, Coish_PRB10, Cywinski_preprint, Witzel_PRL10} with hyperfine interaction to the lattice nuclear spins as the main culprit for
electron spin decoherence.

Decoherence of two-spin states in a coupled double quantum dot is crucial to the operation and scale-up of exchange-based spin quantum computer
architectures.\cite{Loss_PRA98, Kane_Nat98, DiVincenzo_Nat00, Friesen_PRB03, Taylor_NP05, Friesen_PRL07}  Since nuclear spins are the main
sources of single spin decoherence in GaAs quantum dots, where most experimental progress has been made, existing theoretical studies have
focused on the decohering effects of the nuclear spins.\cite{Erlingsson_PRB01, Coish_PRB05,Yang_PRB08,Prada_PRB08} In addition, since exchange
coupling is electrostatic in nature, exchange-coupled electrons are vulnerable to charge noise and other orbital fluctuations that have an
electrical signature.\cite{Hu_PRA00, Coish_PRB05, Hu_PRL06, Culcer_APL09, Roszak_PRB09, Ramon_PRB10}  For example, we have shown how gate noise
\cite{Hu_PRA00} and background charge fluctuations\cite{Hu_PRL06} lead to pure dephasing by introducing noise into exchange splitting of a
double dot.

Electron-phonon interaction is intrinsic to any solid state system,\cite{Mahan,YuCardona} and semiconductor nanostructures are no exception. It
is therefore important to consider the role of electron-phonon interaction in electron spin decoherence.  While electron-phonon interaction is
generally not spin-dependent, it can affect spins when combined with other interactions.  For example, in a single quantum dot, electron-phonon
interaction can assist single-electron spin flip or two-spin transitions in combination with spin-orbit interaction\cite{Fabian_PRL98,
Fabian_PRL99, Khaetskii_PRB01, Golovach_PRL04, Tahan_PRB05, Stano_PRL06, Stano_PRB06, Climente_PRB07, Golovach_PRB08} or hyperfine
interaction.\cite{Erlingsson_PRB01, Erlingsson_PRB02, Prada_PRB08}  In the case of donors, the strongly localized electron wave function and the
resulting lattice strain lead to a direct spin-lattice interaction, so that electron-phonon interaction can cause pure dephasing for a single
spin.\cite{Mozyrsky_PRB02}  For a pair of donors close to each other, the two-electron spin states can be mixed by the hyperfine interaction
with the P nuclear spins, which allows two-spin relaxation via phonon emission.\cite{Borhani_PRB10}

In this work we study the decoherence effects of electron-phonon interaction on two-electron-spin states in semiconductor double quantum dots
(DQD's). Singlet and triplet states are two-spin eigenstates for exchange-coupled electrons in the absence of spin-orbit interaction and
inhomogeneous magnetic fields (otherwise electron-phonon interaction can lead to relaxations between singlet and triplet
states\cite{Climente_PRB07, Golovach_PRB08, Prada_PRB08, Borhani_PRB10}).  These two types of states have different charge distributions because
of their different spin symmetry. We show that this {\it difference} in electron charge density distribution leads to different dressing by the
phonons, without the involvement of the excited states and/or spin-orbit interaction.  This difference in phonon dressing then leads to pure
dephasing between singlet and triplet states. The systems we consider include coupled quantum dots in GaAs and Si, both regarded as promising
candidates for qubits in spin-based quantum information processing.

The paper is organized as follows.  In Sec. II, we introduce the electron-phonon interaction in GaAs and Si.  Combined with knowledge of
two-electron states, we obtain the effective interaction Hamiltonian in the form of a spin-boson model, and clarify the dynamics of two-spin
dephasing. In Sec. III, we present our results, quantifying the time scale of two-spin dephasing in both GaAs and Si, and in both symmetric and
biased double dots, and identifying the most important types of electron-phonon interaction.  Finally, in Sec. IV, we discuss the implications
of our results on spin and exchange-based quantum information processing, and we give our conclusions.

\section{Theoretical Formalism}

\subsection{Electron-phonon interaction in GaAs and Si}

The general electron-phonon interaction Hamiltonian in a semiconductor takes the form \cite{YuCardona}
\begin{equation}
H_{ep} = \sum_{{\bf q},\lambda} M_{\lambda}({\bf q}) \rho({\bf q}) (a_{{\bf q},\lambda} + a_{-{\bf q},\lambda}^\dagger) \,,
\label{eq:eph_general}
\end{equation}
where $a_{{\bf q},\lambda}$ and $a_{-{\bf q},\lambda}^\dagger$ are phonon annihilation and creation operators with lattice momentum ${\bf q}$
and branch index $\lambda$, and $\rho({\bf q})$ is the electron density operator.  For this work we consider the electron interaction with both
acoustic and optical phonons.

For semiconductors with polar characteristics, such as GaAs and InAs, electron-phonon interaction is generally strong, including deformation
potential (DP) interaction and piezoelectric (PE) interaction with acoustic phonons, and polar (PO) interaction with longitudinal optical (LO)
phonons. Deformation potential interaction in GaAs only couples electrons to longitudinal acoustic (LA) phonons,
\begin{equation}
M^{\rm DP}_{\rm GaAs}({\bf q}) = D \left(\frac{\hbar}{\rho V \omega_{\bf q}}\right)^{\frac{1}{2}} \left| {\bf q} \right| \,,
\end{equation}
where $D$ is the deformation constant, $\rho$ is the mass density, $V$ is the volume of the crystal, and $\omega_{\bf q}$ is the angular
frequency of the phonon mode ${\bf q}$.  For GaAs $D = 8.6$ eV and $\rho = 5.3 \times 10^3$ kg/m$^3$.  For piezoelectric interaction in a
zinc-blende lattice,
\begin{equation}
M^{\rm PE}_{\rm GaAs}({\bf q}) = i \left(\frac{\hbar}{\rho V \omega_{\bf q}}\right)^{\frac{1}{2}} 2 e e_{14} \left( \hat{q}_x \hat{q}_y \xi_z +
\hat{q}_y \hat{q}_z \xi_x + \hat{q}_z \hat{q}_x \xi_y \right) \,,
\end{equation}
where $e$ is the elementary electric charge, $e_{14}$ is an elasticity tensor component, $\hat{\xi}$ is the polarization vector, and $\hat{q}$
is the unit vector along ${\bf q}$.  For GaAs $e_{14} = 1.38 \times 10^9$ V/m.  Notice that PE interaction can couple electrons to both LA and
transverse acoustic (TA) phonons.  For polar interaction with LO phonons in bulk polar materials such as GaAs and InAs,
\begin{equation}
M^{\rm PO}_{\rm GaAs}({\bf q}) = \sqrt{\frac{2\pi e^2 \hbar \omega_{LO}}{q^2 V} \left( \frac{1}{\epsilon_\infty} - \frac{1}{\epsilon_0} \right)}
\,,
\end{equation}
where $\epsilon_\infty$ and $\epsilon_0$ are the high- and low-frequency dielectric constants, and $\hbar \omega_{\rm LO}$ is the zone-center LO
phonon energy.  For GaAs, $\epsilon_\infty = 10.89$, $\epsilon_0 = 12.9$, and $\hbar \omega_{\rm LO} = 36.25$ meV.  In a quantum well with well
width $a_z$, where barrier materials have different dielectric constants than the well itself, the LO phonons are confined, so that the LO
phonon wave vectors along the confinement direction can only take discrete values of $q_z = n\pi/a_z$, with $n$ being positive
integers.\cite{Stroscio_book}

For Si, which has a vanishing PE interaction because of the inversion symmetry of its lattice, the DP interaction has similar strength as in
GaAs, and can couple electrons to both acoustic phonon branches.  However, there is no interaction between conduction electrons and optical
phonons in Si.\cite{YuCardona}  The conduction band of bulk Si has a six-fold degeneracy at its bottom,\cite{YuCardona} so that the DP
electron-phonon interaction takes on a more complicated form.\cite{YuCardona}  For an electron in a particular valley along the $\hat{k}$
direction,
\begin{equation}
H_{\rm Si}^{\rm DP} = \Xi_d Tr\{\varepsilon\} + \Xi_u (\hat{k} \cdot \varepsilon \cdot \hat{k})  \,, \label{eq:e-ph-Si}
\end{equation}
where $\Xi_d$ and $\Xi_u$ are the dilation and shear deformation potential constants, and $\varepsilon$ is the strain tensor of the lattice due
to lattice vibrations. For Si, $\Xi_d = 5.0$ eV and $\Xi_u = 8.77$ eV for electrons at the bottom of the conduction
band.\cite{Landolt_Bornstein,YuCardona}  For a two-dimensional quantum dot (in the $xy$-plane) whose electronic ground orbital state involves
only the $z$ and $-z$ valleys,
\begin{eqnarray}
M_{\rm Si}^{\rm DP,LA} ({\bf q}) & = & \Xi_d \left(\frac{\hbar}{\rho V \omega_{\bf q}}\right)^{\frac{1}{2}} \left| {\bf q} \right| \left( 1 +
\frac{\Xi_u}{\Xi_d} \hat{q}_z^2 \right) \,, \\
M_{\rm Si}^{\rm DP,TA} ({\bf q}) & = & \Xi_u \left(\frac{\hbar}{\rho V \omega_{\bf q}}\right)^{\frac{1}{2}} \xi_z q_z \,.
\end{eqnarray}

Having obtained the explicit forms of the electron-phonon interaction Hamiltonians in both GaAs and Si, we can now project them onto a specific
electronic state basis.  Below, we discuss these projections in both symmetric and biased double quantum dots with two electrons.

\subsection{Charge distribution of two electrons in a symmetric double quantum dot}

For two electrons in a DQD, the electron density operator $\rho({\bf q})$ in the general electron-phonon interaction Hamiltonian
Eq.~(\ref{eq:eph_general}) takes the form $\rho({\bf q}) = e^{i{\bf q} \cdot {\bf r}_1} + e^{i{\bf q} \cdot {\bf r}_2}$.\cite{Mahan}  With the
knowledge of electron orbital states, we can calculate the matrix elements of $\rho({\bf q})$.

When two spin qubits are exchange-coupled in an unbiased symmetric DQD, their orbital states are symmetric or anti-symmetric if their spin state
is a singlet ($|\!\!\uparrow\downarrow - \downarrow\uparrow\rangle/\sqrt{2}$) or a triplet ($|\!\!\uparrow\downarrow -
\downarrow\uparrow\rangle/\sqrt{2}$, $|\!\!\uparrow \uparrow \rangle$, $|\!\!\downarrow \downarrow \rangle$). Within the Heitler-London
approximation, the two spatial wave functions can be written as
\begin{eqnarray}
|\psi_S \rangle & = & \frac{1}{\sqrt{2(1+S^2)}} |L(1)R(2)+R(1)L(2) \rangle \,, \nonumber \\
|\psi_{AS} \rangle & = & \frac{1}{\sqrt{2(1-S^2)}} |L(1)R(2)-R(1)L(2) \rangle \,,
\end{eqnarray}
where $L$ and $R$ refer to the ground single-electron orbital states in the two dots, $S=\langle L|R\rangle$ is the overlap integral, and $1$
and $2$ are indices for the two electrons.

Now we can project the electron-phonon interaction into the singlet-triplet Hilbert space.  All three triplet states have the same orbital wave
function and cannot be differentiated by electron-phonon interaction. The Hilbert space of interest is thus only two-dimensional, with the
corresponding basis states $\frac{1}{\sqrt{2(1+S^2)}} |L(1)R(2)+R(1)L(2)\rangle \times \frac{1}{\sqrt{2}} |\!\uparrow\downarrow -
\downarrow\uparrow\rangle$ and $\frac{1}{\sqrt{2(1-S^2)}} |L(1)R(2)-R(1)L(2)\rangle \times \frac{1}{\sqrt{2}} |\!\uparrow\downarrow +
\downarrow\uparrow\rangle$.  Since the Hamiltonian has no spin-dependence, the $2 \times 2$ electron-phonon interaction Hamiltonian is diagonal:
\begin{equation}
H_{\rm eff} = \sum_{{\bf q},\lambda} M_{\lambda}({\bf q}) A_{\phi} \sigma_z (a_{{\bf q},\lambda} + a_{-{\bf q},\lambda}^\dagger)\,,
\label{eq:Hamiltonian_eff}
\end{equation}
where $\sigma_z$ is a Pauli matrix in this two-dimensional two-electron Hilbert space (it is not for single electron spins), and the charge
distribution difference $A_\phi$ is given by
\begin{equation}
A_\phi = \frac{1}{2}[ \langle \psi_{AS}| \rho({\bf q}) |\psi_{AS} \rangle - \langle \psi_{S}| \rho({\bf q}) |\psi_{S} \rangle ] = A_\phi ({\bf
q}_\parallel) f(q_z) \,. \label{eq:A_phi}
\end{equation}
Here $f(q_z)$ is determined by the $z$-direction (growth direction) wave function, and there is no transition between subbands created by
$z$-confinement. For an infinite square well with width $a_z$ and for acoustic phonons (whose wave vectors are not limited by the quantum well
confinement),
\begin{equation}
f(q_z) = \frac{\sin q_z a_z}{q_z a_z} \frac{-\pi^2}{(q_z a_z)^2 - \pi^2} \,.
\end{equation}
For LO phonons, $q_z$ are discrete: $q_z = m\pi/a_z$, with $m = 1, 2, ...$.  In the present calculation there is no intersubband transition, so
that
\begin{equation}
f(q_z = 2n\pi/a_z) = 0 \,,
\end{equation}
while for $q_z = (2n+1)\pi/a_z$,
\begin{equation}
f\left( \frac{(2n+1)\pi}{a_z} \right) = \frac{(-1)^{n+1}}{(n-1/2)(n+1/2)(n+3/2)} \,.
\end{equation}

For a symmetric DQD, the singlet state has larger charge density in between the two dots, while the triplet has larger charge density at the far
ends of the DQD. The resulting difference in charge distribution has a finite electrical quadrupole moment and gives $A_\phi({\bf q}_\parallel)$
its ${\bf q}$-dependence:
\begin{equation}
A_\phi^{Sym} ({\bf q}_\parallel) = \frac{2S^2 e^{-(q_\parallel a)^2/4}}{1-S^4}  \left\{ \cos q_x L - \cosh \left( \frac{q_y a}{2}
\frac{La}{l_B^2} \right) \right\} \,, \label{eq:A_phi_Sym_B}
\end{equation}
where $l_B = \sqrt{\hbar/ eB}$ is the magnetic length for a single electron. At zero external field $A_\phi ({\bf q}_\parallel)$ takes on the
simplified form of
\begin{equation}
A_\phi^{Sym} ({\bf q}_\parallel, B=0) = -\frac{4S^2 e^{-(q_\parallel a)^2/4}}{1-S^4}  \sin^2 \left(\frac{q_x L}{2}\right) \,.
\label{eq:A_phi_Sym_0}
\end{equation}

\subsection{Charge distribution of two electrons in a biased double quantum dot}

In the case of a singlet-triplet qubit,\cite{Petta_Science05} the DQD is biased.  The interdot bias is in the regime where the ground triplet
state remains in the (11) configuration, while the ground singlet state is generally a superposition of the (11) [denoted as S(11)] and (02)
[denoted as S(02)] singlets. In S(11), the two electrons are symmetrically distributed across the two dots.  In S(02), the two electrons are
both in the ground orbital state of the lower-energy dot. S(11) and S(02) are tunnel coupled, and the composition of the ground singlet state
$|{\rm S}\rangle$ depends on the detuning $\epsilon$ between the two singlets in the absence of the tunnel coupling.  Here $\epsilon = 0$ is
defined as the anticrossing point of S(11) and S(02). For negative (positive) $\epsilon$, S(11) [S(02)] has lower energy.
\begin{eqnarray}
|{\rm S}\rangle & = & \alpha |{\rm S}(11)\rangle + \beta |{\rm S}(02)\rangle \\
|{\rm S}(11)\rangle & = & \psi_S \times \frac{1}{\sqrt{2}} |\!\uparrow\downarrow -
\downarrow\uparrow\rangle \\
|{\rm S}(02)\rangle & = & |R(1)R(2)\rangle \times \frac{1}{\sqrt{2}} |\!\uparrow\downarrow - \downarrow\uparrow\rangle \\
|{\rm T}\rangle & = & \psi_{AS} \times \frac{1}{\sqrt{2}} |\!\uparrow\downarrow + \downarrow\uparrow\rangle
\end{eqnarray}
Here for simplicity we assume $\alpha$ and $\beta$ to be real.  Both are functions of the interdot detuning $\epsilon$.  With the $|{\rm
S}\rangle$ and $|{\rm T}\rangle$ states given and the respective charge distributions known, we can calculate the charge distribution difference
for a biased DQD as a function of $\alpha$ and $\beta$:
\begin{widetext}
\begin{eqnarray}
A_\phi^{Biased} ({\bf q}_\parallel) & = & -i \beta^2 e^{-(q_\parallel a)^2/4} \sin q_x L - \alpha \beta \frac{\sqrt{2} S}{\sqrt{1+S^2}}
e^{-(q_\parallel a)^2/4} \left\{ e^{iq_x L}
+ \cosh \left( \frac{q_y a}{2} \frac{La}{l_B^2} \right) \right\} \nonumber \\
& & + \left[ 1 - \frac{\beta^2}{2} (1-S^2) \right] \frac{2S^2}{1-S^4} \ e^{-(q_\parallel a)^2/4} \left\{ \cos q_x L - \cosh \left( \frac{q_y
a}{2} \frac{La}{l_B^2} \right) \right\} \label{eq:A_phi_biased_B}
\end{eqnarray}
Similar to the case of a symmetric DQD, at zero magnetic field the expression of $A_\phi$ is simplified:
\begin{eqnarray}
A_\phi^{Biased} ({\bf q}_\parallel, B=0) & = & - \alpha \beta \frac{2\sqrt{2} S}{\sqrt{1+S^2}} \ e^{-(q_\parallel a)^2/4} \cos^2 \frac{q_x L}{2}
- \left[ 1 - \frac{\beta^2}{2} (1-S^2) \right] \frac{4S^2}{1-S^4} \ e^{-(q_\parallel a)^2/4} \sin^2 \frac{q_x L}{2} \nonumber \\
& & -i \left[ \beta^2 + \alpha \beta \frac{\sqrt{2} S}{\sqrt{1+S^2}} \right] e^{-(q_\parallel a)^2/4} \sin q_x L \label{eq:A_phi_biased_0}
\end{eqnarray}
\end{widetext}

\begin{figure}
\vspace*{0.2in}
\includegraphics[width=3.2in]{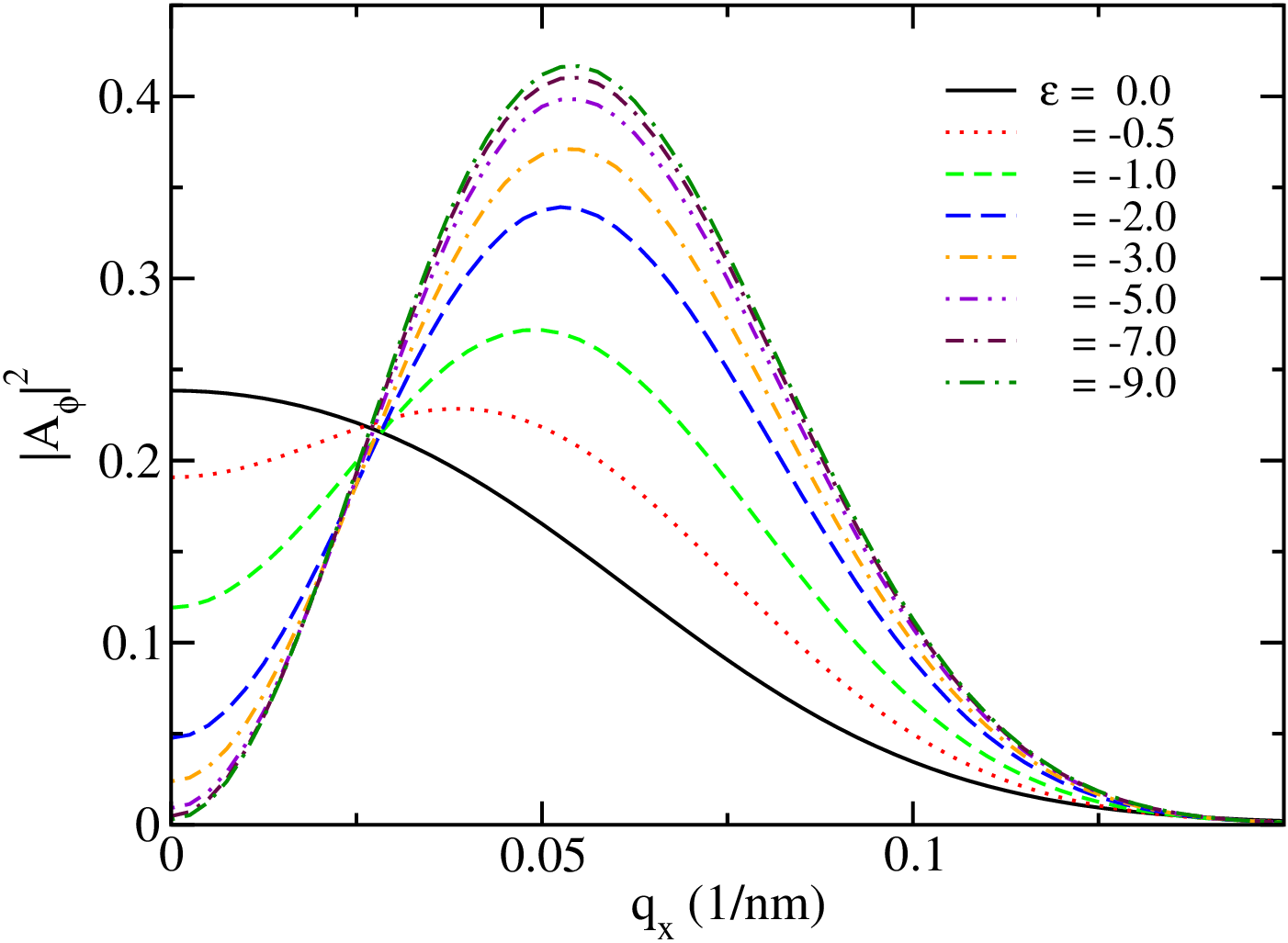}
\caption{(Color online) Charge density difference $A_\phi^{Biased}$ as a function of phonon wave vector $q_x$ and interdot bias $\epsilon$ in a
biased DQD. Notice that as soon as $\epsilon$ moves away from the S(11)-S(02) anticrossing point (where $\epsilon = 0$) toward the negative
bias, $A_\phi$ has similar characteristics in the form of a peak determined by the inter-dot distance $L$.  In this regime the two electrons are
in the (11) configuration, essentially the same as the case of a symmetric DQD.  Close to $\epsilon = 0$, the two-electron singlet state
acquires a (02) component, while the peak of $A_\phi$ shifts toward $q_x = 0$.} \label{fig:A_phi_qx}
\end{figure}

The finite interdot bias, which leads to all the additional terms in $A_\phi$ when $\beta \neq 0$, has some important consequences.  One
distinct feature of Eq.~(\ref{eq:A_phi_biased_0}) is the first term on the right hand side, which does not go to zero when $q_x \rightarrow 0$.
It implies that low-frequency phonons are more efficient in causing dephasing for a biased DQD compared to a symmetric DQD.  In
Fig.~\ref{fig:A_phi_qx} we show $A_\phi^{Biased}$ as a function of $q_x$ for various detuning $\epsilon$.  As discussed above, as $\epsilon$
approaches 0, the S(02) component increases in the ground singlet state, and $A_\phi^{Biased}$ acquires a finite value at $q_x = 0$.  On the
other hand, for $\epsilon \ll 0$ so that $\beta \rightarrow 0$, the biased DQD system approaches the symmetric case, so that $A_\phi^{Biased}
\rightarrow A_\phi^{Sym}$. Furthermore, for the more symmetric DQDs, $A_\phi$ has a peak around $q_x \sim 1/L$, as can be seen from the
functional form of $A_\phi^{Sym}$. Another interesting feature of Eq.~(\ref{eq:A_phi_biased_0}) is the last term on the right hand side, which
apparently does not go to zero when overlap $S \rightarrow 0$ as long as $\beta$ is finite.  This term is again due to the charge distribution
difference between (11) and (02) configurations.  We will explore the consequence of this term at the end of the next section.

\subsection{Two-spin dephasing due to electron-phonon interaction with a dissipative bosonic reservoir}

The effective electron-phonon interaction Hamiltonian of Eq.~(\ref{eq:Hamiltonian_eff}) is a typical spin-boson Hamiltonian that leads to decay
in the off-diagonal element of the $2 \times 2$ density matrix:\cite{Duan_PRA98}
\begin{equation}
\rho_{\rm ST} (t) = \rho_{\rm ST} (0) e^{-B^2(t)}\,,
\end{equation}
where the dephasing factor is positive definite:\cite{Duan_PRA98}
\begin{equation}
B^2(t) = \frac{V}{\pi^3 \hbar^2} \int d^3{\bf q} \frac{|M({\bf q}) A_\phi({\bf q})|^2}{\omega_{\bf q}^2} \sin^2 \frac{\omega_{\bf q} t}{2} \coth
\frac{\hbar \omega_{\bf q}}{k_B T} \,. \label{eq:dephasing_ideal}
\end{equation}

It has long been pointed out that bosonic reservoirs with vanishing density of states at low frequencies do not cause complete decay of the
off-diagonal element of a two-level system density matrix,\cite{Mozyrsky_JSP98, Mozyrsky_PRB02, Krummheuer_PRB02, Vagov_PRB02, Fedichkin_PRA04,
Axt_PRB05, Borri_PRB05, Stavrou_PRB05, Krummheuer_PRB05, Hodgson_PRA10} in other words $B^2(t)$ of Eq.~(\ref{eq:dephasing_ideal}) does not
diverge with time, because bosonic modes with $\omega \rightarrow 0$ determine the long-time behavior for the two-level system. This absence of
complete dephasing can be traced back to the assumptions made when the dephasing formula Eq.~(\ref{eq:dephasing_ideal}) is derived. While it
does account for the fact that the bosonic reservoir is in a thermal equilibrium before getting into contact with the spin,\cite{Duan_PRA98} it
treats the harmonic modes in the bosonic reservoir as completely coherent. However, these harmonic modes, in the present case the phonons, also
belong to an open system, and could lose their coherence to their environments.\cite{Herring_PR54, Ziman_book, Hunyh_PRL06, Rozas_PRL09}  When
relaxations of the bosonic modes are taken into account, it is expected that pure dephasing of the two-level system would eventually become
complete.  For example, in a spin-boson model study of localization, Ref.~\onlinecite{Machnikowski_PRL06} showed how anharmonicity of the
bosonic reservoir, in the specific form of two-phonon scattering (which is one out of four terms in a transformed phonon interaction
Hamiltonian), would lead to complete decoherence of the two-level system considered.  Here we account for phonon relaxation by first deriving
the phonon Langevin equations\cite{Meystre_book} that describes the effects of phonon-reservoir interactions.

As we show in Appendix \ref{App:Langevin}, the Langevin equation for the phonon annihilation operator (in the Heisenberg picture) takes the form
\begin{equation}
\frac{d}{dt}a_{\bf q}(t) = -i\omega_{\bf q} a_{\bf q} (t) - \frac{\gamma_{\bf q}}{2} a_{\bf q} (t) - i\kappa_{\bf q} \sigma_z + F_{\bf q} (t)
e^{-i\omega_{\bf q} t} \,, \label{eq:Langevin}
\end{equation}
where $\omega_{\bf q}$ is the phonon angular frequency, $\gamma_{\bf q}$ is the population relaxation rate of the phonon mode ${\bf q}$,
$\kappa_{\bf q} = M_{\lambda}({\bf q}) A_{\phi}({\bf q})$ is the two-electron-phonon interaction strength from Eq.~(\ref{eq:Hamiltonian_eff}),
$\sigma_z$ is the Pauli operator in the truncated two-electron singlet-triplet space, and $F_{\bf q}(t)$ is a noise operator of the reservoir.
As shown in Appendix \ref{App:Langevin}, the specific forms of $\gamma_{\bf q}$ and $F_{\bf q}(t)$ depend on the reservoir and the
system-reservoir interaction. However, the form of the Langevin equation (\ref{eq:Langevin}) is quite generic. Compared to a dissipationless
phonon mode, now we have the additional second and fourth terms on the right hand side of Eq.~(\ref{eq:Langevin}), representing the dissipation
and fluctuation caused by the coupling to the reservoir. We include these two terms and rederive Eq.~(\ref{eq:dephasing_ideal}) using the
approach adopted in Ref.~\onlinecite{Duan_PRA98}.  The outline is sketched in Appendix \ref{App:Dephasing}.  Now we obtain
\begin{widetext}
\begin{eqnarray}
& & \hspace*{-0.5 in} \rho_{\rm ST} (t) = \rho_{\rm ST} (0) e^{-B_1^2(t)-B^2_2(t)}\,,
\nonumber \\
& & \hspace*{-0.5 in} B_1^2(t) = \frac{V}{4\pi^3 \hbar^2} \int d^3{\bf q} \frac{|M({\bf q}) A_\phi({\bf q})|^2}{\omega_{\bf q}^2 +(\gamma_{\bf
q}/2)^2} \left\{ \frac{\omega_{\bf q}^2 - (\gamma_{\bf q}/2)^2 }{ \omega_{\bf q}^2 + (\gamma_{\bf q}/2)^2 } \left( 1- e^{-\frac{\gamma_{\bf
q}}{2}t} \cos \omega_{\bf q} t \right) - \frac{e^{-\frac{\gamma_{\bf q}}{2}t} \omega_{\bf q} \gamma_{\bf q}/2}{ \omega_{\bf q}^2 + (\gamma_{\bf
q}/2)^2 } \sin \omega_{\bf q} t \right\} \nonumber \\
& & + \frac{V}{2\pi^3 \hbar^2} \int d^3{\bf q} \frac{|M({\bf q}) A_\phi({\bf q})|^2}{\omega_{\bf q}^2 +(\gamma_{\bf q}/2)^2} \left\{ 1 +
e^{-\gamma_{\bf q} t} -2 e^{-\frac{\gamma_{\bf q}}{2}t} \cos \omega_{\bf q} t  \right\} \frac{1}{e^{\frac{\hbar \omega_{\bf q}}{k_B T}} - 1}
\,, \label{eq:dephasing1} \\
& & \hspace*{-0.5 in} B^2_2(t) = \frac{V}{2\pi^3
\hbar^2} \int d^3{\bf q} \frac{|M({\bf q}) A_\phi({\bf q})|^2}{\omega_{\bf q}^2 + (\gamma_{\bf q}/2)^2} \left(\frac{\gamma_{\bf q}}{2} t \right)
= \Gamma_{\rm ST} t \,. \label{eq:dephasing2}
\end{eqnarray}
\end{widetext}
At the limit that phonon decay rate $\gamma_{\bf q} \rightarrow 0$, $B_1^2(t) \rightarrow B^2 (t)$ while $B_2^2(t) \rightarrow 0$.  For finite
$\gamma_{\bf q}$, corresponding to a dissipative phonon reservoir, we obtain an additional exponential decay of the off-diagonal density matrix
element in Eq.~(\ref{eq:dephasing2}) compared to the non-dissipative reservoir result of Eq.~(\ref{eq:dephasing_ideal}).  The rate of this
exponential decay $\Gamma_{\rm ST}$ is proportional to the phonon decay rate $\gamma_{\bf q}$ integrated over the phonon modes.  Notice that
$\Gamma_{\rm ST}$ does not explicitly contain the thermal factor $\coth \frac{\hbar \omega_{\bf q}}{k_B T}$ that describes the thermal
occupation of the phonon modes. This is because $B_2^2 (t)$ comes from phonons decaying into their reservoirs, when phonons themselves are
regarded as coherent bosons, while temperature information of the reservoirs for the phonons is contained in the decay rate $\gamma_{\bf q}$ and
the noise operator $F_{\bf q} (t)$.

\section{Results}

The main questions we would like to answer in this work are as follows: Is electron-phonon interaction an important decoherence channel for spin
qubits in semiconductor quantum dots?  Under what condition is it important?  How do different substrate materials (GaAs and Si) compare to each
other? And how do different qubit architectures compare with each other?  Below we show our results that provide the answers.

\subsection{Symmetric double dot}

Let us first examine the dynamical behaviors of the dephasing factors $B^2(t)$ and dephasing rate $\Gamma_{\rm ST}$ due to electron-phonon
interaction when the double quantum dot is unbiased.

\begin{figure}
\includegraphics[width=3.2in]{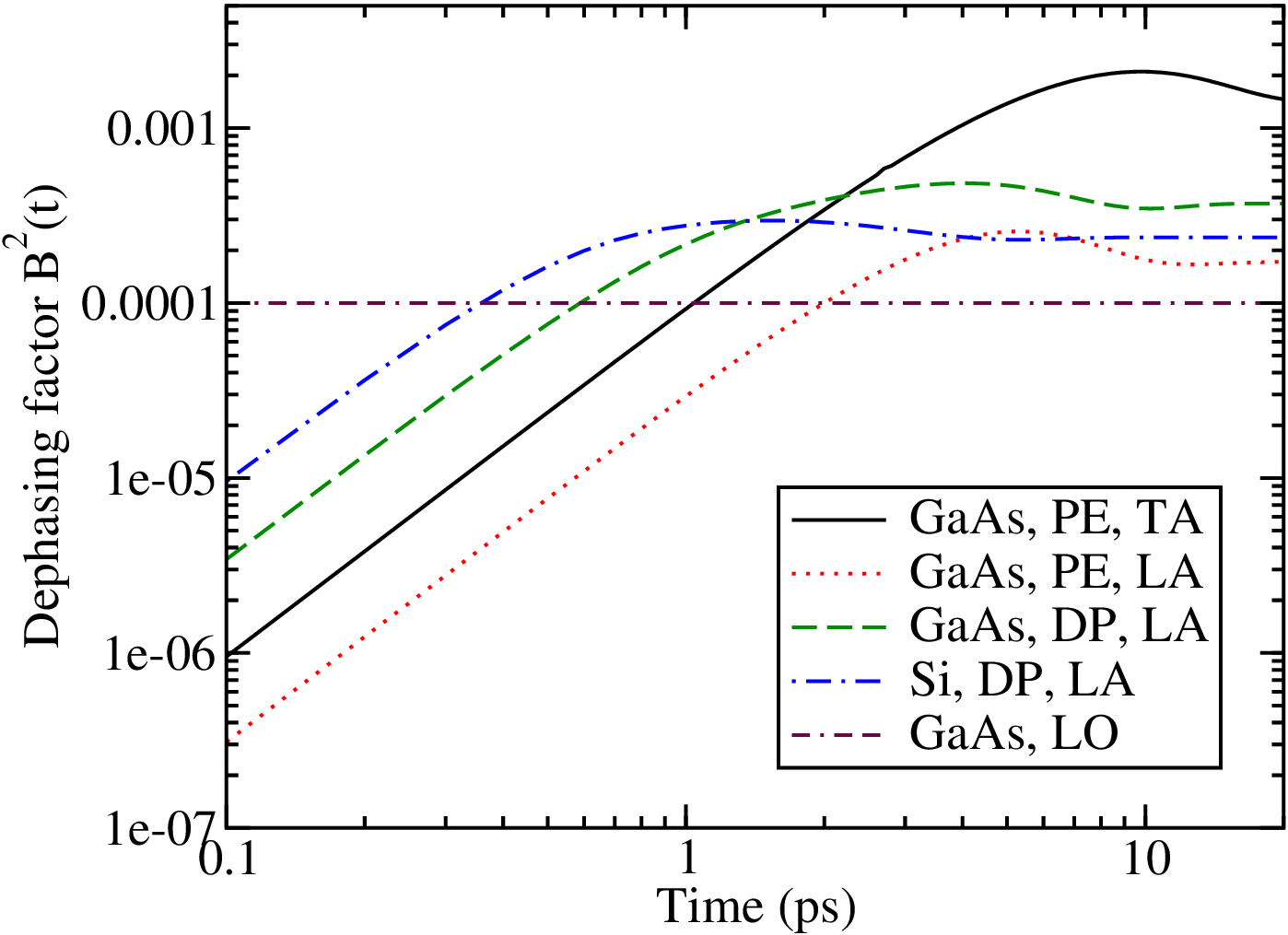}
\caption{ (Color online) Two-spin dephasing in a symmetric double quantum dot induced by a non-dissipative phonon reservoir.  All the curves are
for double quantum dots with an interdot separation of 40 nm and single dot orbital radius of 20 nm. More specifically, the black solid line is
for PE interaction with TA phonons in GaAs, the red dotted line is for PE coupling to LA phonons in GaAs, the green dashed line is for DP
coupling to LA phonons in GaAs; the blue dot-dashed curve is for DP coupling to LA phonons in Si; and the maroon dot-dashed-dashed horizontal
line represents the dephasing magnitude for polar interaction with LO phonons in GaAs.  The dephasing here is given by $B^2(t \rightarrow
\infty)$} \label{fig:dephasing_nondissipative}
\end{figure}

In Fig.~\ref{fig:dephasing_nondissipative} we show the typical behavior of the dephasing factor $B^2(t)$ in the absence of phonon decay for
various types of electron-phonon interactions in GaAs and Si.  There are two interesting features shared by all the curves for acoustic phonons
in Fig.~\ref{fig:dephasing_nondissipative}.  At very short times ($t \ll 1$ ps), the increase of $B^2(t)$ is quadratic, which originates from
Taylor expansion of the $\sin^2 \omega_{\bf q} t /2$ factor in the integrand at the small-$t$ limit.  At long times, all the curves saturate,
which means that dephasing does not increase with time anymore, so that it corresponds more to a finite loss of contrast than the conventional
complete decay of off-diagonal density matrix elements.  The transition between the quadratic increase and the saturation happens between 1 and
10 ps for double dots with a dot separation of 40 nm and a single-dot radius of 20 nm because this time is essentially determined by the
interdot distance divided by the speed of sound ($5 \sim 8 \times 10^3$ m/s in Si and $3 \sim 5 \times 10^3$ m/s in GaAs): 40 nm / c $\sim$ 10
ps.  The saturation time for Si is shorter because Si has a larger speed of sound.  The dephasing factor due to DP interaction with TA phonons
in Si is two orders of magnitude smaller than that due to LA phonons, and is not plotted in Fig.~\ref{fig:dephasing_nondissipative}.

Mathematically, the long-time saturation can be understood by writing $2\sin^2 \omega_{\bf q} t /2$ as $1 - \cos \omega_{\bf q} t = 1-\cos
(cqt)$.  Since the acoustic phonon spectrum is continuous and nonsingular, the cosine term leads to a vanishing contribution to the integral at
large times, which leaves the dephasing factor determined by a constant integral that is independent of time.  Physically, this saturation is
due to the fact that long-time dephasing is determined by the low-frequency part of the spectrum of the bosonic reservoir, while the phonon
density of states vanishes quadratically at low frequency.  In other words, non-dissipative acoustic phonons simply form an inefficient
dephasing reservoir as compared to other charge fluctuation reservoirs such as fluctuating charge traps, which have a $1/f$ spectral density.
This incomplete dephasing has been observed theoretically in a variety of calculations related to phonons, in the studies of general spin-boson
decoherence behaviors, charge and spin coherence of single electrons, and exciton coherence.\cite{Mozyrsky_JSP98, Mozyrsky_PRB02,
Krummheuer_PRB02, Vagov_PRB02, Fedichkin_PRA04, Axt_PRB05, Borri_PRB05, Stavrou_PRB05, Krummheuer_PRB05, Hodgson_PRA10}

For electron interaction with optical phonons in GaAs, the dephasing factor $B^2(t)$ takes on a particularly simple form because the optical
phonon dispersion at the zone center is flat.  Take $\omega_{\bf q} \approx \omega_{LO}$, we obtain
\begin{eqnarray}
B^2_{LO}(t) & = & \frac{V}{\pi^3 \hbar^2} \int d^3{\bf q} \frac{|M({\bf q}) A_\phi({\bf q})|^2}{\omega_{\bf q}^2} \sin^2 \frac{\omega_{\bf q}
t}{2} \coth \frac{\hbar \omega_{\bf q}}{k_B T} \nonumber \\
& = & \frac{2 e^2}{\pi^2 \hbar \omega_{LO}} \left( \frac{1}{\epsilon_\infty} - \frac{1}{\epsilon_0} \right) \coth \frac{\hbar \omega_{LO}}{k_B
T}
\sin^2 \frac{\omega_{LO} t}{2} \nonumber \\
& & \times \int d^3{\bf q} \frac{|A_\phi({\bf q})|^2}{q^2} \nonumber \\
& = & b^2_{LO} \sin^2 \frac{\omega_{LO} t}{2} \,,
\end{eqnarray}
which is a sinusoidal function of time.  For a GaAs double dot with a single-dot radius of $a=20$ nm, and $L/a$ in the range of 1 and 2, the
coefficient $b^2_{LO}$ for the sinusoidal function ranges between $10^{-4}$ and $10^{-9}$.

\begin{figure}
\includegraphics[width=3.2in]{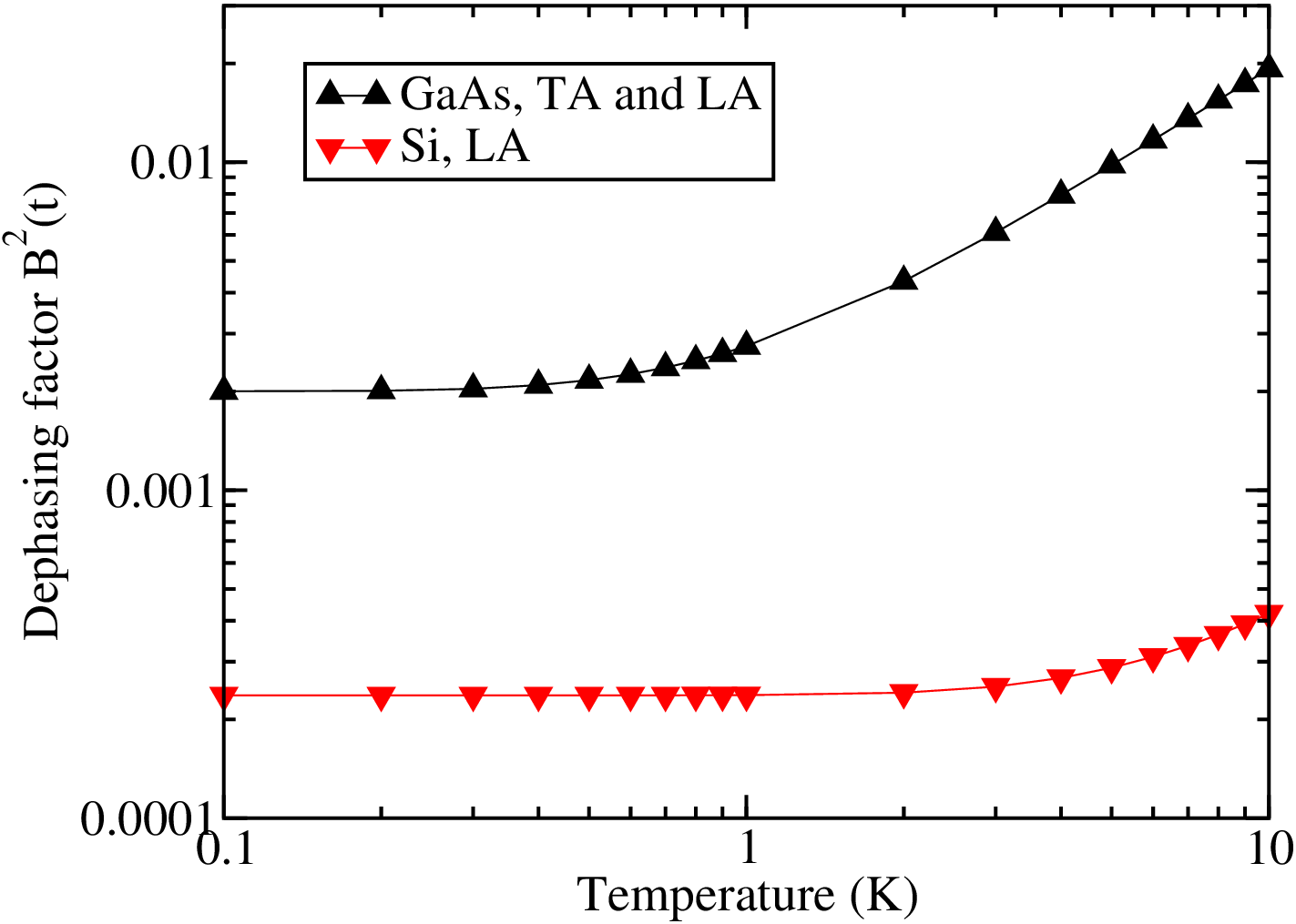}
\caption{ (Color online) Phonon induced two-spin dephasing rate as a function of acoustic phonon temperature in a symmetric double dot for both
GaAs and Si. The single dot wave function radius for all the data is 20 nm. } \label{fig:dephasing_T}
\end{figure}

As we discussed in the previous section, the absolute value of the saturated dephasing, as long as it is small ($\ll 1$), is not an important
parameter by itself because dephasing will eventually become complete due to phonon relaxation.  However, the relative magnitudes of the
saturated dephasing shown in Fig.~\ref{fig:dephasing_nondissipative} do give a qualitative sense of the relative importance of various types of
electron-acoustic-phonon interactions. Specifically, in GaAs PE coupling to TA phonons produces the strongest dephasing effect, while in Si DP
coupling to LA phonons is the most important.  In addition, as indicated in Eq.(\ref{eq:dephasing_ideal}), $B^2(t)$ does have a strong
temperature dependence as well.  At higher temperatures more acoustic phonon modes contribute to dephasing, so that $B^2(t)$ can eventually
become an $O(1)$ quantity and dephasing can be considered complete. In Fig.~\ref{fig:dephasing_T} we plot the temperature dependence of the
saturated $B^2(t)$ for GaAs and Si quantum dots.  At temperatures above 1 K dephasing increases with temperature almost linearly.  On the other
hand, optical phonon induced dephasing does not have a pronounced temperature dependence even at $T = 10$ K because $\hbar \omega_{LO} \sim 36$
meV is much larger than $k_B T$ at low temperatures.

When phonon decay is included, the most important effect is the added exponential dephasing $e^{-\Gamma_{\rm ST} t}$, upon which we will focus
in the rest of this study. To calculate $\Gamma_{\rm ST}$, we need to first identify the ${\bf q}$-dependence of the phonon relaxation rate
$\gamma_{\bf q}$. Qualitatively, lower energy (lower frequency $\omega_{\bf q}$) acoustic phonons have to decay slower (smaller $\gamma_{\bf
q}$). For example, when phonon decay is due to anharmonicity, or more specifically the third order process of one phonon splitting into two, the
phonon decay rate could vary as $q^n$ with $n$ between 1 and 4 depending on the lattice symmetry and phonon branches.\cite{Herring_PR54,
Ziman_book} In the case of a phonon cavity, it is not clear how the $Q$-factor would vary with the phonon wave vector, although with
$\gamma_{\bf q} \propto D(\omega_{\bf q})$ [where $D(\omega_{\bf q})$ is the phonon reservoir density of states] one could expect $\gamma_{\bf
q} \propto q^2$. In the following, we calculate the acoustic-phonon-induced dephasing rate assuming that $\gamma_{\bf q} = \gamma_0 q^n$, with
$n$ taking the value of 2 or 3. Taking $Q = 10^3$ for a TA phonon with an energy of $0.1$ meV, we obtain $\gamma_0 = 10^8$ 1/s. For LA phonons,
which have higher energies than TA phonons with the same $q$, we take $\gamma_0 = 10^9$ 1/s. This is an arbitrary choice that is used to reflect
the fact that LA phonons generally have shorter lifetimes than TA phonons.\cite{Ziman_book} For LO phonons we assume a constant relaxation time
of 10 ps for all modes. The rationale here is that LO phonons have a flat dispersion, so that they should have a near constant relaxation rate
near the Brillouin zone center. The LO phonons also have a very short lifetime because of their large energy. Zone center LO phonons have been
measured to have a lifetime of 7 ps.\cite{Linde_PRL80, Ridley_PRB91}

\begin{figure}
\includegraphics[width=3.2in]{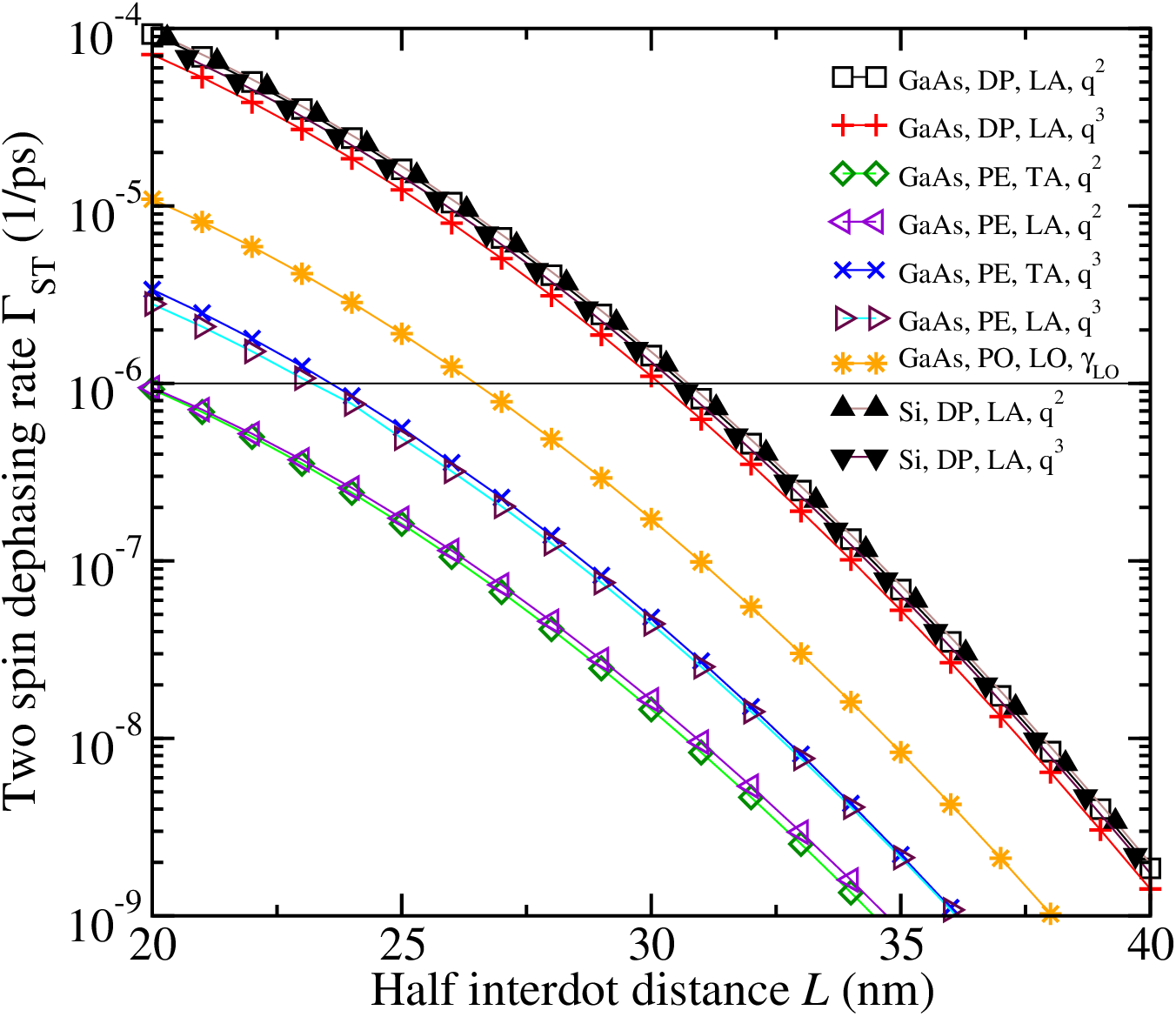}
\caption{(Color online) Phonon-induced two-spin dephasing rate as a function of the interdot separation for a GaAs and a Si symmetric double
quantum dot.  The horizontal line is drawn at a dephasing time of 1 $\mu$s, approximately the decoherence times measured in
Refs.~\onlinecite{Petta_Science05,Koppens_PRL08}.  The legends for the data sets have the following format: type of materials (GaAs or Si), type
of interaction (DP, PE, or PO), type of phonons involved (LA, TA, or LO), and the $q$-dependence of $\gamma_{\bf q}$ ($q^2$, $q^3$, or constant
$\gamma_{LO}$).} \label{fig:dephasing_dissipative_L}
\end{figure}

In Fig.~\ref{fig:dephasing_dissipative_L}, we plot the phonon-induced two-spin dephasing rate $\Gamma_{\rm ST}$ in symmetric DQDs in both GaAs
and Si as functions of the half interdot distance $L$.  The radius of the single dot electron wave function is 20 nm for this figure and all the
following figures. The strong dependence on $L$ for all data sets originates from the fact that the charge distribution difference between the
two-electron singlet and triplet states in a symmetric DQD is directly dependent on interdot wave function overlap: $\Gamma_{\rm ST} \propto
[4S^2/(1-S^2)]^2$, so that the smaller the overlap, the smaller the difference in charge distribution, and the smaller the phonon-induced
dephasing.  Based on the data given in Fig.~\ref{fig:dephasing_dissipative_L}, phonon-induced dephasing is not an important decoherence
mechanism when $L/a > 2$ in a symmetric DQD.

An important feature of Fig.~\ref{fig:dephasing_dissipative_L} is that DP coupling to LA phonons is the most important dephasing channel for
GaAs, and produces about the same magnitude of dephasing in Si.  In GaAs, dephasing due to DP coupling is about one order of magnitude larger
than that by PO coupling to LO phonons, and almost two orders of magnitude larger than that due to PE coupling to both LA and TA phonons.  This
fact is somewhat surprising because in Fig.~\ref{fig:dephasing_nondissipative} it is clear that PE coupling to TA phonons is by far the most
important decoherence channel. However, notice that in the present calculation of $\Gamma_{\rm ST}$ the acoustic phonon decay increases rapidly
as phonon energy increases, so that the contributions from higher-energy phonons are much more important in the calculation of $\Gamma_{\rm ST}$
than in $B^2(t)$.  This tilt toward higher-energy phonons strongly favors DP coupling over PE couplings because of the factor of $q$ difference
in the electron-phonon coupling matrix element.  The similar values for dephasing for GaAs and Si within the DP mechanism is more of a
coincidence: they have similar values in DP coupling strength, mass density, and speed of sound, and we chose the same $\gamma_0$ for both
materials, although Si has the extra contribution from the shear DP constant $\Xi_u$. Based on the results presented in this figure,
phonon-induced dephasing is an essentially equivalent decoherence mechanism for Si and GaAs.

Another interesting aspects of Fig.~\ref{fig:dephasing_dissipative_L} is that LO phonons turn out to be a strong source of dephasing for the
two-spin states in GaAs, even though they have very high energy in GaAs ($\sim$ 36 meV).  This somewhat surprising result originates from the
facts that LO phonons have a diverging density of states at the zone center (as compared to the vanishing density of states for the acoustic
phonons) and a very fast relaxation rate (experimentally measured at 7 ps\cite{Linde_PRL80}), and that GaAs has a reasonably strong polar
interaction strength. In Si, the conduction electrons do not interact with optical phonons, therefore this decoherence channel is completely
removed.

For DP and PE interactions, the different $q$-dependence of the phonon relaxation rate $\gamma_{\bf q}$ leads to quite different results in the
two-spin dephasing rate $\Gamma_{\rm ST}$. With DP interaction, $\Gamma_{\rm ST}$ is not very sensitive to the exponent $n$, and increasing $n$
in $\gamma_{\bf q} \sim q^n$ leads to a slight decrease of $\Gamma_{\rm ST}$.  On the other hand, for PE interaction, increasing $n$ leads to an
approximately three-fold increase of $\Gamma_{\rm ST}$.  The change of the exponent $n$ leads to a shift of the dominant ${\bf q}$ region that
contributes to dephasing. For PE interaction, increasing $n$ from 2 to 3 moves the dominant contribution to larger $q$ phonons, which have a
larger density of states, leading to an increase in $\Gamma_{\rm ST}$.  For DP interaction, the dominant contribution already comes from the $qa
\sim 1$ region, where changing $n$ does not have much of an effect.

\begin{figure}
\includegraphics[width=3.2in]{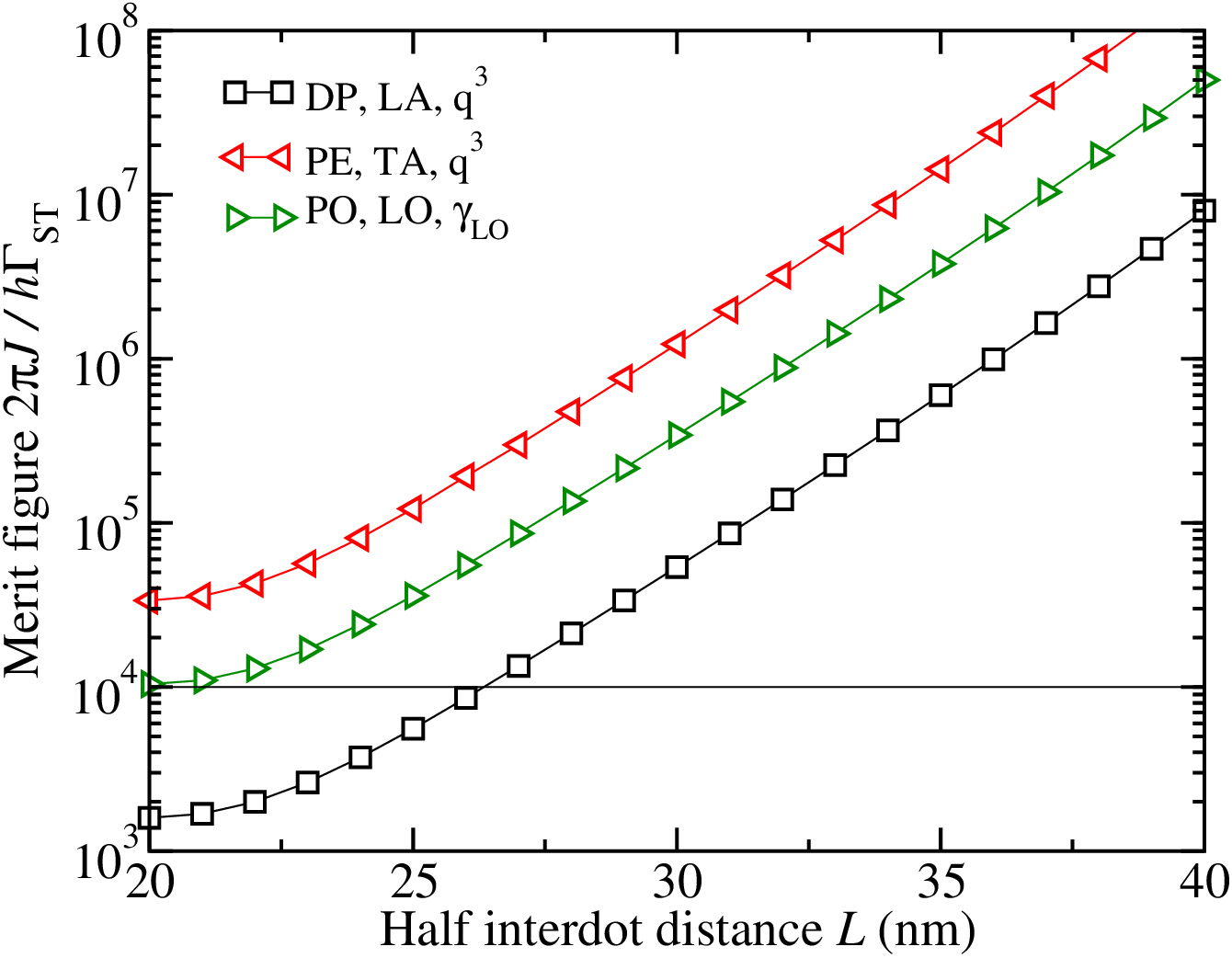}
\caption{(Color online) Merit figure based on phonon-induced dephasing of two-spin states in a symmetric GaAs double dot as a function of half
interdot distance. We draw a line at $10^4$ as the nominal threshold for fault tolerant quantum computation.  Therefore the double dot (with
single-dot wave function radius at 20 nm) should be kept apart further than 60 nm.  The legend format is similar to in
Fig.~\ref{fig:dephasing_dissipative_L} (without the first item for materials as all data here are for GaAs): type of coupling, type of phonon,
and $q$-dependence of $\gamma_{\bf q}$.} \label{fig:merit}
\end{figure}

In Fig.~\ref{fig:merit}, we plot the two-spin merit figure $\cal{M}$ as a function of the interdot distance for double dots in GaAs. Here the
merit figure is defined as the ratio between a typical exchange gate time given by $\hbar/J$ ($J$ is the exchange splitting) and the two-spin
decay time given by $1/\Gamma_{\rm ST}$: ${\cal M} = J/ \hbar \Gamma_{\rm ST}$.  The exchange splitting $J$ is calculated within the
Heitler-London model with a quartic confinement potential \cite{Burkard_PRB99}.  The increase of the merit figure at larger inter-dot distance
reflects the fact that the exchange splitting and the phonon-induced dephasing have a different dependence on the interdot overlap integral $S$:
$J \sim S^2$, while $\Gamma_{\rm ST} \sim S^4$. The results shown in this figure indicate that for a two-dot exchange gate to operate with a low
error rate, a slower operation with a smaller interdot overlap is preferable with regard to phonon-induced dephasing, and fault-tolerant
two-qubit operations should be achievable for pretty strongly coupled dots, with $L/a \gtrsim 1.5$.  We do not have any data for Si DQDs in this
figure. Calculating exchange interaction in a Si double dot requires much more sophisticated quantum chemical approaches than a simple
Heitler-London approximation \cite{Hu_PRA00, Scarola_PRA05, Li_PRB10} because in Si the interaction effect is stronger compared to GaAs (larger
effective mass and smaller dielectric constant), so that the Heitler-London approximation does not adequately account for the two-electron
correlation.  For the current evaluation, it is sufficient to point out that Fig.~\ref{fig:dephasing_dissipative_L} above indicates that
phonon-induced dephasing is about the same order of magnitude in Si as in GaAs, while exchange coupling should only be somewhat smaller than in
GaAs. Therefore overall the merit figure should remain about the same when moving from GaAs to Si.

\subsection{Biased double dot}

For a biased DQD, the main question is whether the admixture from the S(02) singlet state and the resulting dipole coupling would lead to
significantly increased dephasing.  Interestingly, the bias not only directly affects the value of $A_\phi({\bf q}_\parallel)$, but also its
functional form.  In Fig.~\ref{fig:A_phi_qx} we have shown how $A_\phi (q_x, q_y = 0)$ depends on $q_x$ for various interdot biases $\epsilon$.
It is clear from that figure that the peak of $A_\phi$ shifts toward $q_x = 0$, while the peak height decreases, as the interdot bias shifts
from the (11) toward the (02) regime.  Furthermore, Eq.~(\ref{eq:A_phi_biased_0}) indicates that as soon as $\beta \neq 0$, there is a mixing of
S(11) and S(02) states, so that $A_\phi^{Biased}$ acquires a nonvanishing component [1st term on the right hand side of
Eq.~(\ref{eq:A_phi_biased_0})] as $q_x \rightarrow 0$, leading to an increase in the phonon-induced dephasing.

\begin{figure}
\includegraphics[width=3.2in]{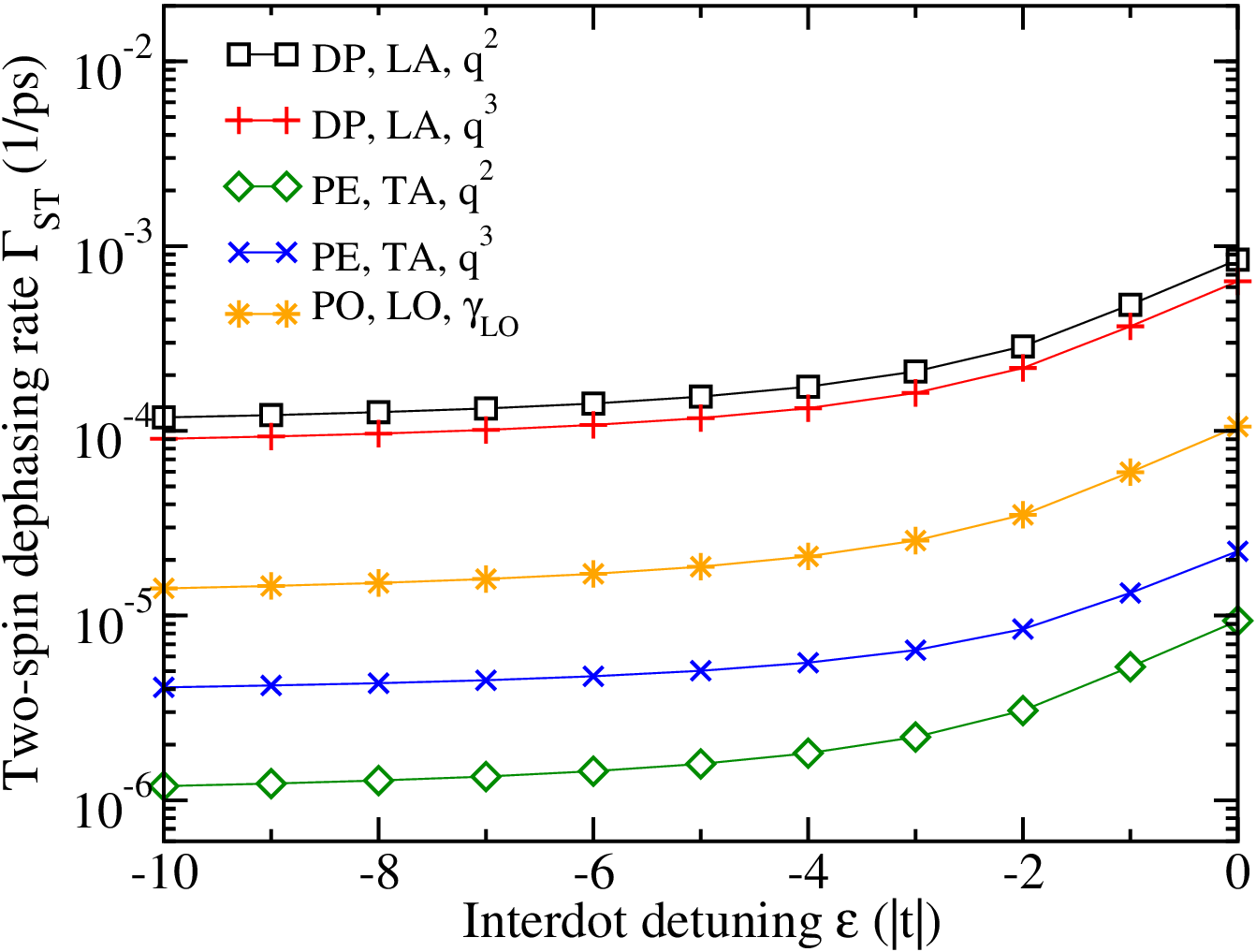}
\caption{ (Color online) Phonon-induced two-spin dephasing rate as a function of the interdot bias $\epsilon$ for a GaAs double quantum dot with
piezoelectric coupling to transverse phonons.  The double dot are separated by 40 nm and the single-dot wave function radius is 20 nm, so that
we are at the strong coupling limit.  At large negative bias the relaxation rate approaches a value of about $10^{-6}$ 1/ps, or $T_\phi$ of
about 1 $\mu$s.  As the bias increases toward the S(11)-S(02) anticrossing, the dephasing rate increases so that at $\epsilon = 0$, $T_\psi \sim
100$ns.} \label{fig:dephasing_epsilon}
\end{figure}

Now we can calculate the two-spin dephasing rate $\Gamma_{\rm ST}$ for any voltage bias between the dots.  Figure~\ref{fig:dephasing_epsilon}
shows $\Gamma_{\rm ST}$ as a function of dimensionless interdot bias $\epsilon$. As $\epsilon$ becomes increasingly negative, the biased DQD
states approach those of a symmetric DQD, and $\Gamma_{\rm ST}$ approaches the value given in Fig.~\ref{fig:dephasing_dissipative_L}. On the
other hand, as $\epsilon$ increases toward positive bias, the ground singlet state has a larger S(02) component, and $A_\phi$ a larger dipolar
contribution, so that $\Gamma_{\rm ST}$ increases.  At $\epsilon = 0$, $\Gamma_{\rm ST}$ is dominated by the dipolar contribution from the
S(11)-S(02) mixing, and is about ten times larger than in a symmetric DQD, where $A_\phi$ is determined by a quadrupolar charge distribution
difference between S(11) singlet and T(11) triplet states.

\begin{figure}
\includegraphics[width=3.2in]{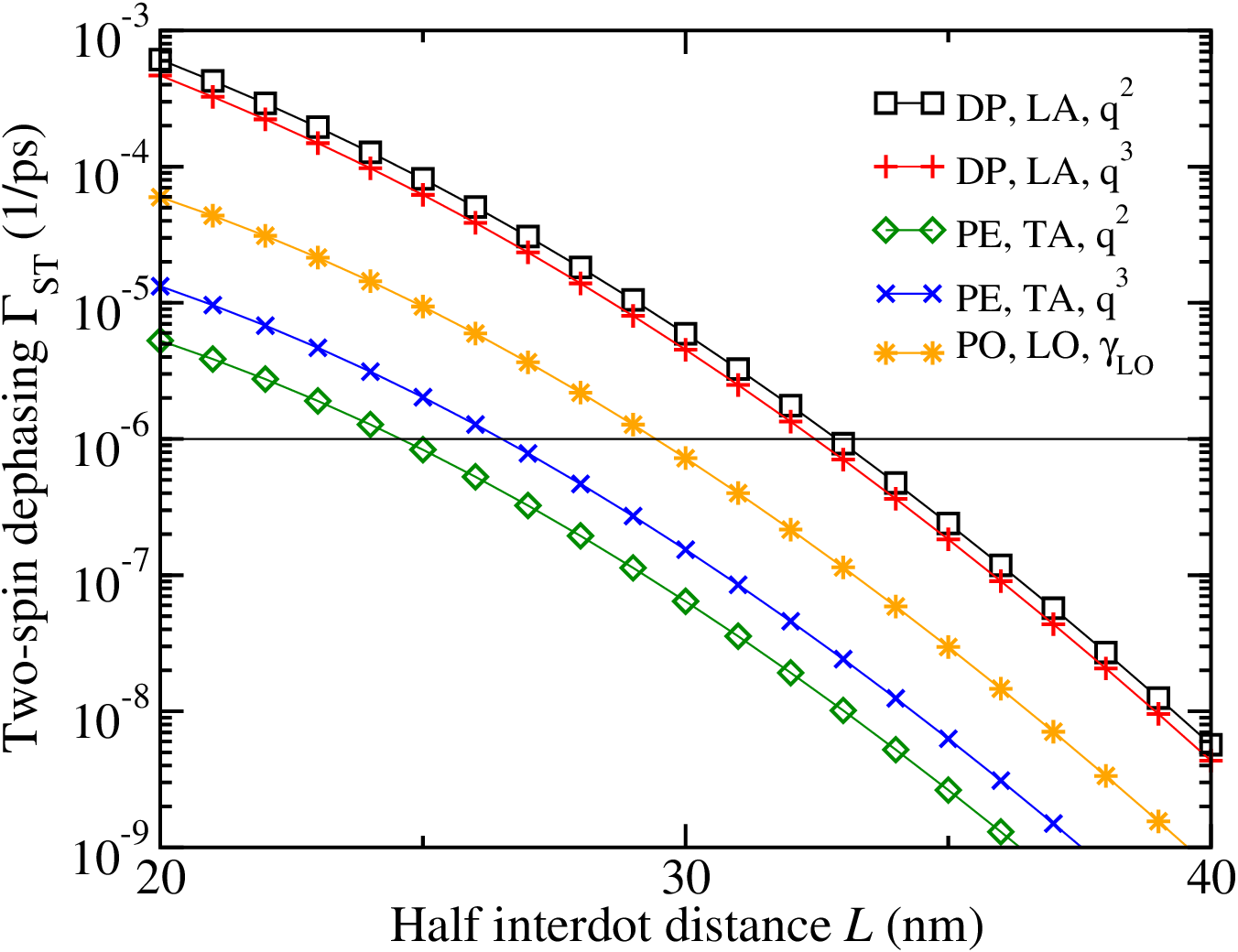}
\caption{ (Color online) Phonon-induced two-spin dephasing rate as a function of the interdot distance $L$ for a biased GaAs double quantum dot
with DP coupling to LA phonons, PO coupling to LO phonons, and PE coupling to TA phonons.  The format of the legend is the type of coupling, the
type of phonon, and the $q$-dependence of the phonon relaxation rate $\gamma_{\bf q}$.  The single-dot wave function radius is 20 nm, and the
interdot bias is $\epsilon = -1$ at $L = 20$ nm.} \label{fig:dephasing_L_biased}
\end{figure}

As indicated in Fig.~\ref{fig:dephasing_dissipative_L}, for symmetric DQDs phonon-induced decoherence becomes much less important at larger $L$
because of the overlap factor $S$ in the charge difference $A_\phi$. In the case of a biased DQD, when the DQD has a vanishing overlap,
\begin{equation}
A_\phi^{Biased} (B=0,S=0) = -i \beta^2 e^{-(q_\parallel a)^2/4} \sin q_x L \,,\label{eq:A_phi_S=0}
\end{equation}
which does not seem to depend on the interdot overlap.  This term leads to dephasing between T(11) and S(02) states, which clearly have
different charge distributions.  However, if overlap $S$ vanishes because the interdot distance $L$ increases, phonon-induced dephasing will not
saturate to a constant, as Eq.~(\ref{eq:A_phi_S=0}) seems to indicate, because $\beta$ depends on $L$ as well.  Recall that the tunnel coupling
$t$ of the S(11) and S(02) singlet states is $t = \langle {\rm S}(11)|H|{\rm S}(02)\rangle \propto S = e^{-L^2/a^2}$.  When $L$ increases, $t$
decreases as $t = t_0 \ e^{-(L^2 - L_0^2)/a^2}$ for two coupled parabolic dots, where $t_0$ is the tunnel coupling at $L_0$.  For a fixed
interdot bias $\epsilon$ [recall that $\epsilon = 0$ corresponds to the S(11)-S(02) crossing point, where $\beta^2 = 0.5$], the DQD moves
farther from the anticrossing point as $t$ gets smaller, leading to a decreasing $\beta$ [which is the weight of the higher-energy singlet; for
negative $\epsilon$, it is the weight of the S(02) state]. In Fig.~\ref{fig:dephasing_L_biased} we plot the dephasing rate for a biased DQD as a
function of the interdot separation. The figure shows the same rapid decrease of dephasing for all types of phonons as $L$ increases, similar to
the situation in symmetric DQDs.  Indeed, putting $\beta \propto S$ into Eq.~(\ref{eq:A_phi_biased_0}), it is clear that $A_\phi^{biased}
\propto S^2$, which is the same overlap-dependence as in the case of a symmetric DQD.

The results in Figs.~\ref{fig:dephasing_epsilon} and \ref{fig:dephasing_L_biased} show that in a biased DQD phonon-induced dephasing approaches
the symmetric DQD limit at large negative bias, and increases monotonically as interdot bias $\epsilon$ increases.  However, one can always
reduce this dephasing by increasing the interdot distance and reducing the wave-function overlap.  Furthermore, for larger negative biases
[deeper into the (11) regime, with smaller exchange splitting $J = t/2\epsilon$], $\beta^2$ is smaller. When $|\epsilon| \gg 1$, $\beta^2 \sim
(1/\epsilon)^2$ while $J \propto 1/\epsilon$. Therefore, there should exist a regime where the dephasing rate $\Gamma_{\rm ST}$ is much smaller
than exchange splitting, so that fault-tolerant exchange gates can be performed. For example, if we choose $L/a = 2$ with $a = 20$ nm,
$1/\Gamma_{\rm ST} \sim 100~\mu$s even at $\epsilon = -1$, according to Fig.~\ref{fig:dephasing_L_biased}.  At such an interdot separation, $|t|
\sim 10 \mu$eV, so that $J \sim 1~\mu$eV for $\epsilon = -10$, with a gate time in the order of a nanosecond, leading to a merit figure of $\sim
10^5$.

\section{Discussions and Conclusions}

Based on our results presented in this study, phonon induced two-spin dephasing in both symmetric DQD's and biased DQD's can be strongly
suppressed by reducing the double-dot tunnel coupling.  The strong overlap dependence of the dephasing rate dictates that phonon-induced
dephasing is only important when the double dot is tightly coupled.  Dephasing for a biased double dot does increase with bias because of the
admixture of the S(02) state in the singlet ground state, which introduces electric dipole coupling into phonon-induced decoherence.  Therefore,
phonon-induced two-spin dephasing is generally stronger in biased DQD's, such as in the case of a singlet-triplet qubit.

Phonon-induced two-spin dephasing studied in this paper is related to the different dressing that singlet and triplet electronic states
experience through interaction with the phonons.  When phonons themselves decohere, this spin dephasing channel leads to true complete
decoherence.  On the other hand, ensemble average over phonon modes (while each evolve coherently) only leads to a finite degree of dephasing. A
legitimate question here is whether this part of the dephasing (due to phonon population average) would disappear if we consider dressed
electron spin states, especially considering that this finite dephasing generally saturates in the order of 10 ps, much faster than the electron
spin initialization and manipulation processes in quantum dots. Mathematically, the answer to this question may very well be ``yes'', as long as
one can identify the energies of the dressed states precisely. But to answer this question with confidence, one needs to clarify how the
energies of the spin states are measured, and how electron-phonon interaction may be incorporated in the description of measurement.  In the
current generation devices, spin detection is achieved through charge sensing in the spin-blockade regime,\cite{Ono_Science02} which is
insensitive to phonons, so that the phonon-induced dephasing due to ensemble averaging cannot be removed. Ultimately, though, this question is
moot because phonons do relax and are not coherent forever.

The phonon-induced pure dephasing mechanism we consider here originates from the charge distribution difference between states that have
different spatial symmetry, and involves no real or virtual phonon emission or absorption.  It is different from another mechanism of
phonon-induced dephasing studied in Ref.~\onlinecite{Roszak_PRB09}, which is based on different level distribution of electron singlet and
triplet states and involves virtual emission and absorption of phonons.

In conclusion, we have studied phonon-induced pure dephasing between two-electron singlet and triplet spin states in a semiconductor double
quantum dot.  We find that this pure dephasing is important for tightly coupled double dots, but is strongly suppressed when the double dot
separation increases, so that at relatively large dot separations ($L/a > 2$) fault-tolerant exchange gates can be realized.  A biased double
dot has stronger dephasing compared to a symmetric double dot with the same dot parameters due to the mixing of (11) and (02) singlet states and
the resulting finite electric-dipole coupling.  We have quantified two-spin dephasing in both GaAs and Si double dots, finding that deformation
potential coupling to LA phonons is the most important dephasing mechanism in both materials, and produces about the same magnitude dephasing in
both materials.  We also find that the LO phonon makes a non-negligible contribution to dephasing in GaAs because of the very fast optical
phonon relaxation. Overall, phonon-induced two-spin dephasing is an equivalent decoherence mechanism for Si and GaAs, is stronger in a biased
double dot than in a symmetric double dot, and can be suppressed by reducing the interdot overlap of the electron wave functions.

This work is supported by NSA and LPS through ARO, and DARPA QuEST through AFOSR.  We also gratefully acknowledge the hospitality and financial
support of the Joint Quantum Institute at the University of Maryland and Kavli Institute of Theoretical Physics at the University of California
at Santa Barbara, where part of this work was performed. We have benefited greatly from useful discussions with Peter Yu, Luming Duan, Hendrik
Bluhm, Sankar Das Sarma, Guy Ramon, and Susan Coppersmith.

\appendix

\section{Phonon relaxation}
\label{App:Langevin}

We can model phonon relaxation by assuming that each phonon mode couples to a continuum of bosonic modes.  This is identical to the description
of a cavity photon mode coupling to a continuum.\cite{Meystre_book} Such a model is grounded in the development of phonon cavities in
semiconductor heterostructures,\cite{Hunyh_PRL06,Rozas_PRL09} but more importantly, it presents a clear physical picture about phonon
relaxation.  As we discuss below, the exact form of phonon-reservoir interaction does not change the general features of phonon relaxation.  The
Hamiltonian for a phonon mode and its reservoirs takes the form
\begin{eqnarray}
H & = & H_s + H_r + H_{int} \nonumber \\
H_s & = & \hbar \omega_{\bf k} a_{\bf k}^\dagger a_{\bf k} \nonumber \\
H_r & = & \sum_{\bf q} \hbar \omega_{\bf q} b_{\bf q}^\dagger b_{\bf q} \nonumber \\
H_{int} & = & \hbar \sum_{\bf q} \left[ g({\bf k,q}) a_{\bf k}^\dagger b_{\bf q} + g^*({\bf k,q}) a_{\bf k} b_{\bf q}^\dagger \right] \,,
\nonumber
\end{eqnarray}
where $a_{\bf k}$ is the phonon annihilation operator, $b_{\bf q}$ is the annihilation operator of the bosonic modes in the reservoir, and
$g({\bf k,q})$ is the coupling strength between the phonon mode and the reservoir modes.  With this interaction with the reservoir, the Langevin
equation for the phonon annihilation operator (in the Heisenberg picture) takes the form (as is obtained in the discussion of cavity photon
decay in any number of quantum optics books, e.g., Ref.~\onlinecite{Meystre_book}, Chap. 14, Sec. 14.3)
\begin{eqnarray}
\frac{d}{dt}a_{\bf k}(t) & = & -i\omega_{\bf k} a_{\bf k} (t) - \frac{\gamma_{\bf k}}{2} a_{\bf k} (t) + F_{\bf k} (t) e^{-i\omega_{\bf k} t}
\label{eq:Langevin1} \\
F_{\bf k}(t) & = & -i \sum_{\bf q} g({\bf k,q}) \tilde{b}_{\bf q}(t_0) e^{-i(\omega_{\bf q} - \omega_{\bf k})t} \label{eq:noise1} \\
\gamma_{\bf k} & = & 2\pi D(\omega_{\bf k}) \left| g(\omega_{\bf k}) \right|^2 \,, \label{eq:gamma1}
\end{eqnarray}
where $\tilde{b}_{\bf q} = b_{\bf q} e^{i\omega_{\bf q} t}$ is the slowly varying amplitude of the reservoir bosonic operator.  Notice that the
noise operator $F_{\bf k}(t)$ here is assumed to be independent of the initial time $t_0$. In the definition of the decay rate $\gamma_{\bf k}$
the sum over reservoir mode ${\bf q}$ has been replaced by an integration over energy, with $D(\omega)$ being the density of state for the
reservoir, and $g({\bf k,q})$ is assumed to be smooth in any narrow energy range so that it can be replaced by $g(\omega)$.

In a realistic system, phonon relaxation often originates from phonon anharmonicity.   The lowest-order anharmonic interaction has each phonon
mode coupled to pairs of other modes in the phonon reservoir (three-phonon processes).  We can choose any particular phonon mode as the system
and consider its dynamics coupling to the rest of the phonon modes,
\begin{eqnarray}
H & = & H_0 + H_{int} \nonumber \\
H_0 & = & \hbar \omega_{\bf q} a_{\bf q}^\dagger a_{\bf q} \nonumber \\
H_{int} & = & \hbar \sum_{\bf q,q'} \left[ g({\bf q,q'}) a_{\bf q}^\dagger a_{\bf q'} a_{\bf q-q'} + g^*({\bf q,q'}) a_{\bf q} a_{\bf
q'}^\dagger a_{\bf q-q'}^\dagger \right] \,, \nonumber
\end{eqnarray}
For simplicity and clarity, we have limited ourselves to normal processes for phonon relaxation.\cite{Stroscio_book}  Following the same
standard procedure as above,\cite{Meystre_book} we first obtain the Heisenberg equations for the operator $a_{\bf q}$, then formally solve its
equation of motion, and finally put the results back into the equation of motion for $a_{\bf q}$.  The only significant difference here is that
each phonon mode plays both the role of a higher energy phonon relaxing to two lower energy ones, and the role of a lower energy mode to which a
higher-energy phonon can decay.  Keeping only the lowest order contributions, we obtain
\begin{widetext}
\begin{eqnarray}
& & \hspace*{-0.35in} \frac{d}{dt}a_{\bf q}(t) = -i\omega_{\bf q} a_{\bf q} (t) - \frac{\gamma_{\bf q}}{2} a_{\bf q} (t) + F_{\bf q} (t)
e^{-i\omega_{\bf q} t}
\label{eq:Langevin2} \\
& & \hspace*{-0.35in} F_{\bf q}(t) = -i \sum_{\bf q} g^*({\bf q,q'}) \tilde{a}_{\bf q'}(t_0) \tilde{a}_{\bf q-q'}(t_0) e^{-i(\omega_{\bf q'} +
\omega_{\bf q-q'} - \omega_{\bf q})t} -i \sum_{\bf q} g({\bf q',q}) \tilde{a}_{\bf q'}(t_0) \tilde{a}_{\bf q'-q}^\dagger (t_0) e^{-i(\omega_{\bf
q'} - \omega_{\bf q'-q} - \omega_{\bf q})t} \label{eq:noise2} \\
& & \hspace*{-0.35in} \gamma_{\bf q} = \sum_{\bf q'} 2\left| g({\bf q,q'}) \right|^2 (1 + n_{\bf q'} + n_{\bf q-q'}) \delta(\omega_{\bf q} -
\omega_{\bf q'} - \omega_{{\bf q} - {\bf q'}}) + \sum_{\bf q'} 2\left| g({\bf q',q}) \right|^2 (n_{\bf q'-q} - n_{\bf q'}) \delta(\omega_{\bf
q'} - \omega_{\bf q} - \omega_{{\bf q'} - {\bf q}})\,, \label{eq:gamma2}
\end{eqnarray}
\end{widetext}
where $\tilde{a}_{\bf q} = a_{\bf q} e^{i\omega_{\bf q} t}$ is again the slowly varying amplitude of the phonon operator. In obtaining the
Langevin equation (\ref{eq:Langevin2}) for the phonon operator $a_{\bf q}$ we have assumed that the population of the relevant reservoir phonon
modes is not strongly perturbed by the phonon interaction, which allows us to factor out the phonon operators from the time integrals and
replace number operator $a_{\bf q}^\dagger a_{\bf q}$ by a constant thermal phonon population number $n_{\bf q}$. Notice that the expression for
$\gamma_{\bf q}$ contains two parts.  The first part is due to two-phonon emission and is proportional to $1 + n_{\bf q'} + n_{\bf q-q'}$,
including both spontaneous and stimulated emissions.  The expression is consistent with the population decay rate obtained using Fermi's Golden
Rule, which is also proportional to $1 + n_{\bf q'} + n_{\bf q-q'}$.\cite{Stroscio_book}  The second part of $\gamma_{\bf q}$ is due to the
absorption process where the ${\bf q}$-mode phonon and a $({\bf q' - q})$-mode phonon are converted into a higher-energy ${\bf q'}$-mode phonon.
This contribution is always positive at thermal equilibrium since $n_{\bf q'} < n_{\bf q'-q}$.  Another interesting feature is that while in the
case of cavity phonon decay $\gamma_{\bf q}$ is only determined by the reservoir density of states and coupling strength [see
Eq.~(\ref{eq:gamma1})], here it is also influenced by the thermal occupation of the reservoir phonon modes.

An important result above is that Eqs.~(\ref{eq:Langevin2}) and (\ref{eq:Langevin1}) are of exactly the same form. This is not really
surprising: both coupling to a bosonic reservoir through a cavity mirror and anharmonic interaction with other bosonic modes lead to relaxation,
as has been clearly demonstrated in numerous experiments and clarified in many theoretical studies.\cite{Meystre_book, Stroscio_book}  We can
therefore use Eq.~(\ref{eq:Langevin}) to describe dissipative phonon dynamics, with appropriately chosen phonon relaxation rate $\gamma_{\bf q}$
and noise operator $F_{\bf q}(t)$.

\section{Phonon induced spin dephasing}
\label{App:Dephasing}

The effective spin-boson Hamiltonian for the two-electron-phonon interaction is given by Eq.~(\ref{eq:Hamiltonian_eff}):
\begin{equation}
H_{eff} = \sum_{{\bf q},\lambda} M_{\lambda}({\bf q}) A_{\phi}({\bf q}) \sigma_z (a_{{\bf q},\lambda} + a_{-{\bf q},\lambda}^\dagger)\,.
\end{equation}
This is a pure dephasing Hamiltonian: the interaction with phonons does not change the two-electron eigenstates and there is no relaxation due
to phonons.  We derive the phonon induced dephasing following the approach presented in Ref.~\onlinecite{Duan_PRA98}. The first step is to
obtain the complete phonon Langevin equation (the Heisenberg equation in Ref.~\onlinecite{Duan_PRA98} because phonon relaxation is not
considered there), which contains both the terms as derived in Appendix \ref{App:Langevin} and a term from the electron-phonon interaction
Hamiltonian above.  The phonon Langevin equation now takes the form
\begin{equation}
\frac{d}{dt}a_{\bf q}(t) = -i\omega_{\bf q} a_{\bf q} (t) - \frac{\gamma_{\bf q}}{2} a_{\bf q} (t) -i \kappa_{\bf q} \sigma_z + F_{\bf q} (t)
e^{-i\omega_{\bf q} t} \,,
\end{equation}
where $\kappa_{\bf q} = M_{\lambda}({\bf q}) A_{\phi}({\bf q})$ for any particular phonon branch.  This differential equation for the phonon
operators can be formally solved.  The solution is
\begin{widetext}
\begin{equation}
a_{\bf q}(t) = a_{\bf q}(0) e^{-(\frac{\gamma_{\bf q}}{2} + i\omega_{\bf q})t} + i \kappa_{\bf q}^* \sigma_z \frac{e^{-(\frac{\gamma_{\bf q}}{2}
+i\omega_{\bf q})t} - 1}{i\omega_{\bf q} + \frac{\gamma_{\bf q}}{2}} + \int_0^t dt' F_{\bf q}(t') e^{-(\frac{\gamma_{\bf q}}{2} + i\omega_{\bf
q})(t-t')} \,. \label{eq:Langevin_solution}
\end{equation}
\end{widetext}
This solution is formal because the noise operator $F_{\bf q} (t)$ contains a sum over all the phonon operators themselves.  However, when we
consider a phonon reservoir that is not driven far away from its equilibrium, which is the case studied in this work, the noise operator would
then essentially be a mean field average over all the phonon modes and can be treated as independent from individual phonon mode properties.
Therefore, to a good approximation, Eq.~(\ref{eq:Langevin_solution}) gives the phonon evolution in the presence of electron-phonon and
phonon-phonon or general phonon-reservoir interactions.

The time evolution of the phonon operators allows us to solve for the total density matrix of the electron-phonon system, because the
electron-phonon interaction is diagonal (i.e. pure dephasing spin-boson interaction), as discussed in Ref.~\onlinecite{Duan_PRA98}.  The phonon
operators can then be traced out using a coherent-state representation for the phonon modes, assuming an initial thermal equilibrium
distribution.\cite{Duan_PRA98}  One subtle point in the derivation of the total density operator is that there is a sign difference between the
von Neumann equation for density operators and the Heisenberg equation for regular operators.  While one can expand a density operator in terms
of regular operators, $\rho (0) = \sum_n a_n \hat{A}_n(0)$, the time evolution is reversed: $\rho (t) = \sum_n a_n \hat{A}_n(-t)$.  Thus we need
to calculate $a_{\bf q} (-t)$ before calculating the density operator.  It is important to note here that for $a_{\bf q} (-t)$, the phonon decay
term $e^{-\gamma_{\bf q} t/2}$ remains a decay term, because $\gamma_{\bf q}$ comes from a second-order perturbation calculation and is not
affected by the change of time direction:
\begin{widetext}
\begin{equation}
a_{\bf q}(-t) = a_{\bf q}(0) e^{-(\frac{\gamma_{\bf q}}{2} - i\omega_{\bf q})t} - i \kappa_{\bf q}^* \sigma_z \frac{e^{-(\frac{\gamma_{\bf
q}}{2} - i\omega_{\bf q})t} - 1}{-i\omega_{\bf q} + \frac{\gamma_{\bf q}}{2}} - \int_0^t dt' F_{\bf q}(-t') e^{-(\frac{\gamma_{\bf q}}{2} -
i\omega_{\bf q})(t-t')} \,. \label{eq:Langevin_solution}
\end{equation}
\end{widetext}

The key difference between the present calculation and the one in Ref.~\onlinecite{Duan_PRA98} is that here the solution to the phonon operator
contains the additional terms for phonon decay and noises.  The decay term is fully integrated in our calculation and leads to the exponential
decay of the off-diagonal matrix element of the two-spin density matrix.  The noise term, as we discussed before, represents an overall mean
field noise that is independent of individual phonon modes or the electrons. With the assumption that the noise term commutes with the system
density operator (i.e. assuming the noise is classical), its effect would be to induce a phase shift in the evolution of the reduced electron
density operator but not decay.

\end{document}